# NEGATIVE DIFFERENTIAL RESISTANCE IN GRAPHENE-BASED BALLISTIC FIELD-EFFECT-TRANSISTOR WITH OBLIQUE TOP GATE

Mircea Dragoman[1*], Adrian Dinescu[1], and Daniela Dragoman[2]

[1]National Institute for Research and Development in Microtechnology (IMT), P.O. Box 38-160, 023573 Bucharest, Romania,

[2]Univ. Bucharest, Physics Faculty, P.O. Box MG-11, 077125 Bucharest, Romania

**Negative differential resistance (NDR) with room temperature peak-valley-ratio of 8 has been observed in a ballistic field-effect-transistor (FET) based on graphene, having an oblique top gate. Graphene FETs with a top gate inclination angle of $45^0$ and a drain-source distance of 400 nm were fabricated on a chip cut from a 4 inch graphene wafer grown by CVD. From the 60 measured devices, NDR was observed only in the regions where the CVD graphene displays the Raman signature of defectless monolayers. In other specific positions on the wafer, where graphene quality was not high enough and the Raman signature indicated the presence of defects, the ballistic character of transport is lost and the graphene FETs display nonlinear drain-voltage dependences tuned by the top and back gate voltage.**

Negative differential resistance (NDR) is observed in many nanostructures, including molecular devices [1], ZnO nanobelts [2], room temperature quantum dots with a very high peak-valley-ratio (PVR) of 80 in dark state and 2240 when illuminated with a 20 W lamp [3], GaN nanowires [4] and even graphene solar cells [5]. The above references, while not exhaustive, are an indication of the fact that there is a wealth of physical mechanisms, especially in nanostructures, which can produce NDR.

Graphene field-effect-transistors (FETs) are other systems in which NDR is present. This phenomenon has been observed, for example, in graphene monolayer/few atom layer thick boron nitride/graphene monolayer heterostructures [6], in which the PVR is high but the current is small, limited to 50-80 nA. It has also been evidenced in non-ballistic graphene FETs [7], with a relative small PVR but currents as high as 1 mA for a gate length of 1 $\mu$m. Yet again, NDR with good PVR and a maximum drain current of 30 $\mu$A at a gate length of 5 $\mu$m [8] has been demonstrated in dual-gate (top- and back-gate) graphene FETs with applications in non-Boolenean logic circuits and their architecture. In this context, it is worth mentioning that, since many years, logic circuits and non-Boolean logic devices are implemented using resonant tunneling diodes based on semiconducting heterostructures, which also show NDR dependences in the current-voltage characteristics [9]. However, NDR is a natural property of graphene under certain bias conditions due to the symmetry of its electronic band structure [8], and atomistic simulations for the ballistic regime show that the transmission has strong minima at moderate energies for equal and low values of the drain and top voltages.

A significant progress in this area would be the fabrication of a ballistic FET-like transistor and the experimental evidence of NDR in this device. This is the aim of this paper. The practical interest in NDR is not limited to its applications in logic circuits. NDR is a basic building block of any oscillator, which contains in addition a LC circuit or antenna, and its strong nonlinear characteristic is a key element in nonlinear electronics, for instance in multipliers, mixers or even detectors of very high frequencies. In these last cases, the cutoff frequency of the NDR device becomes a paramount issue. None of the NDR devices mentioned above has a cutoff frequency exceeding 10 GHz, while the need for nonlinear electronics starts at 60-100 GHz and finishes at 4-6 THz. A ballistic FET-like graphene device could fulfill this prerequisite of very high cutoff frequencies.

**Ballistic FET on graphene monolayer with oblique gate: fabrication and measurement results.**

The creation of a bandgap in graphene monolayers involves rather difficult technological procedures (see the references in Ref. 8, for instance), almost all of them destroying the physical properties of the graphene monolayer. On the contrary, in ballistic graphene devices an oblique gate modulates the transmission between the two electrodes: drain and source. Because of the lack of an energy bandgap, the transmission is constant and equal to unity in ballistic graphene devices with gates normal to the direction of electron propagation, irrespective of gate biases, whereas oblique gates at an angle of $45^o$, for instance, induce a transmission bandgap of about 0.3 eV without significant degradation of the physical properties of graphene monolayers (see [10]-[12] for the theoretical analysis of charge carrier transmission through oblique gates in graphene monolayers). Such oblique gates, acting as barriers [10], could generate NDR [11] in a ballistic transport regime and could be used to implement ultrafast Schottky diodes in graphene monolayers [12]. In particular, the NDR in ballistic graphene devices with oblique gates occurs due to a sudden drop to zero of the transmission of charge carriers over a certain drain voltage $V_D$ range, caused by the impossibility of charge carrier propagation through the gated region of the bandgapless graphene monolayer. This decrease in transmission leads to a corresponding minimum in $I_D$ versus $V_D$ characteristic, the drain current $I_D$ being determined by the transmission coefficient via the Landauer formula. A bias applied on the oblique gate of the ballistic graphene FET can shift the NDR region of the $I_D$-$V_D$ dependence and can affect the PVR of the device.

We have fabricated FETs with oblique gates on a chip cut from CVD graphene transferred on a doped 4 inch silicon wafer, having already deposited a 285 nm thin film of silicon dioxide on top of it. The scanning electron microscopy (SEM) photo of the FET is displayed in Fig. 1, the drain and source electrodes being denoted by D and S, respectively. The graphene wafer was provided by Graphene Supermarket. The graphene rectangle in Fig. 1 was shaped using electron beam lithography (Raith e_Line) for patterning and reactive ion etching for removal of the carbon layer. The rectangle was covered with a positive resist PMMA 950k A2 that was exposed at 30 kV and 300 µC/cm$^2$, and then developed in MIBK:IPA (1:3). We used a positive electron resist instead of a negative one to avoid irradiation of the graphene channel with electrons, because electron irradiation degrades the transport properties of graphene. The metallic contacts (5 nm Cr/100 nm Au) were fabricated by e-beam lithography and lift-off, a highly directional e-beam evaporator being used for metal deposition in order to facilitate the lift-off process. Then, the structure was covered with HSQ, which was lithographically shaped (e-beam irradiation at 30 kV and 800 µC/cm$^2$) to create the gate dielectric. The HSQ, with an electrical permittivity around 3, was used many times as gate dielectric in graphene FETs and it is among the most used resists in graphene e-beam processing because it does not harm significantly graphene. The thickness of the HSQ was about 50 nm. On top of the gate insulating layer, the tilted gate consisting of a Cr/Au electrode with a thickness of 45 nm (5 nm Cr and 40 nm Au) was fabricated by e-beam lithography and lift-off. The gate length is 40 nm, at the state-of-art of graphene FETs [13].

The graphene CVD transferred on the 4 inch Si/SiO$_2$ wafer displays two main distinct features: (i) about 70 % of the monolayer graphene area is practically defectless, (ii) 30 % of the graphene area contains defects (cracks, wrinkles, multilayer graphene). The Raman signature of the defectless graphene monolayer in Fig. 2(a) contains the G band at 1593.9 cm$^{-1}$ and the 2D peak located at 2651.7 cm$^{-1}$, the ratio between the 2D and G bands being about 2. No defect band D appears in this Raman signal. On the contrary, the Raman spectrum in Fig. 2(b), taken in a graphene area with defects, shows a peak associated to the D band, which indicates that the area under investigation is no longer an ideal graphene monolayer. In principle, it is a very tedious task to perform a Raman map of such a big chip, but a couple of Raman measurements must be done to locate the monolayer areas.The Raman spectra were obtained with a Labram Hr800 spectrometer and a laser excitation wavelength of 633 nm.

All 60 graphene FETs fabricated on the graphene chip were measured in DC with the help of a Keithley 4200 SCS equipment with low noise amplifiers at its outputs. The DC probes connected to the Keithley 4200 were placed together with the probe station in a Faraday cage provided by Keithley. This system also allows the application of a back gate voltage up to 100 V via an interlock circuit. The graphene FET chip connected with the electrical probes for measurements is shown in Fig. 3. All measurements were done at room temperature with the source electrode connected to the ground.

In the monolayer graphene regions we have seen NDR in all measured graphene FETs, whereas in the graphene area with defects the NDR behavior was not observed, the ballistic transport was lost, and the devices displayed FET-like $I_D$-$V_D$ dependences that could be tuned by top and back gate voltages.  From the 60 graphene FETs, 8 did not worked at all.

As an example, a NDR with a tunable PVR depending on the top-gate voltage is represented in Fig. 4, where the $I_D$-$V_D$ curves were measured at top gates of $V_{TG}$ = 0 V (red dashed line), 0.5 V (blue dotted line) and 1 V (solid magenta line). The PVR ratio is enhanced by the gate voltage, as predicted in Ref. [11] in the case of a ballistic graphene FET with an oblique gate. The top gate voltage changes the Fermi level in graphene and thus modifies the shape of the NDR. For a top gate voltage of 1 V, the maximum drain current attains 105 µA at a drain voltage of 1.5 V, and decreases to about 12 µA at $V_D = 2$ V. The PVR for this top gate voltage value is about 8.75. On the contrary, when no gate voltage is applied the maximum and minimum $I_D$ values are about 95 µA and 12 µA, respectively, the corresponding PVR being 7.9. This NDR behavior was retrieved in many graphene FETs with oblique gates having PVRs in the 6-8 range, tunable by top gate voltages. The cutoff frequency of such a ballistic and tunable NDR device depends only on the transit time of carriers between drain and source. Thus, the cutoff frequency of the ballistic NDR in Fig. 1, with a drain-source distance of 400 nm and a Fermi velocity of $10^6$ m/s, is in the THz range.

However, the NDR behavior is lost in a non-ballistic transport regime. A typical $I_D$-$V_D$ dependence in this case, shown in Fig. 5 for several top and gate voltages, looks different compared to a typical graphene FET. Unlike in the latter device, in which the $I_D$-$V_D$ dependences are linear, in our graphene FETs they are clearly nonlinear. Because the graphene-Cr/Au contact is ohmic (Cr has the same workfunction as graphene monolayer, i.e. –4.5 eV), the nonlinearity can be attributed to the oblique gate. The various $I_D$-$V_D$ curves in Fig. 5 correspond to: $V_{TG} = -2$ V (green solid line), $V_{TG} = -1$ V (magenta dotted line), $V_{TG} = 0$ V(red dotted line), $V_{TG} = 1$ V (blue dotted line), $V_{TG} = 2$ V (solid blue line), $V_{TG} = 2$ V and $V_{BG} = 40$ V (solid magenta line), $V_{TG} = 2$ V and $V_{BG} = 50$ V (green dotted line), and $V_{TG} = 2$ V and $V_{BG} = 60$ V (solid red line). The different $I_D$-$V_D$ characteristics in the graphene region with defects compared to those in defectless monolayer graphene can be explained by the fact that scattering on defects and impurities randomize the propagation direction of charge carriers, changing also their energy if the scattering is inelastic. Because the position and width of the transmission gap that gives rise to NDR depends on the energy and propagation direction of charge carriers, it follows that the NDR disappears due to scattering-induced averaging, even if the phase of charge carrier wavefunction is preserved at scattering. From Fig. 5 it can be seen that the $I_D$-$V_D$ characteristic is influenced by the top and gate voltages, the graphene FET being practically blocked for $V_{TG} = 2$ V and $V_{BG} = 60$ V, the corresponding characteristic being superimposed on the horizontal axis. At such a high back gate voltage the spatial charge of the very large number of charge carriers screens the applied source-drain voltage. Another effect of the non-ballistic transport regime is that the maxim drain current is now about 40 µA – i.e. it is reduced with more than 50% compared to the ballistic device in Fig. 4.

**Conclusions**

We have fabricated tens of graphene FETs having a tilted top gate on a CVD grown graphene monolayer chip. A large part of them, situated in the defect-free areas of the chip, are working in the ballistic regime and display NDR with a large PVR, having cutoff frequencies in the THz region. The other devices, situated in the area where defects are present, lose the NDR behavior, but display nonlinear $I_D$-$V_D$ dependences tuned by top and back gate voltages. In the non-ballistic propagation regime, due to the scattering-induced lengthening of the charge carriers' path in th device, the graphene FETs have a lower cutoff than those working in the ballistic regime, but could still be used in various RF applications.

*Acknowledgements* We thank the European Commission for the financial support via the FP 7 NANO RF (grant agreement 318352). We also thank to eng. Adrian Albu for his help in the DC measurements of the graphene FET devices.

FIGURE CAPTIONS

Fig. 1 SEM photo of the FET based on a graphene monolayer, with an oblique gate.

Fig. 2 (a) Raman signature of the defectless graphene monolayer, and (b) Raman signature of the graphene region with defects.

Fig. 3 The graphene FET chip under preparation to be measured on Keithley 4200 SCS.

Fig. 4 NDR behavior in the ballistic regime of the graphene FET with oblique gate at $V_{TG} = 0$ V (red dashed line), 0.5 V (blue dotted line) and 1 V (solid red line).

Fig. 5 Drain current-drain voltage dependences of the graphene FET at various top and back gate voltages: $V_{TG} = -2$ V (green solid line), $V_{TG} = -1$ V (magenta dotted line), $V_{TG} = 0$ V(red dotted line), $V_{TG} = 1$ V (blue dotted line), $V_{TG} = 2$ V (solid blue line), $V_{TG} = 2$ V and $V_{BG} = 40$ V (solid magenta line), $V_{TG} = 2$ V and $V_{BG} = 50$ V (green dotted line), and $V_{TG} = 2$ V and $V_{BG} = 60$ V (solid red line).

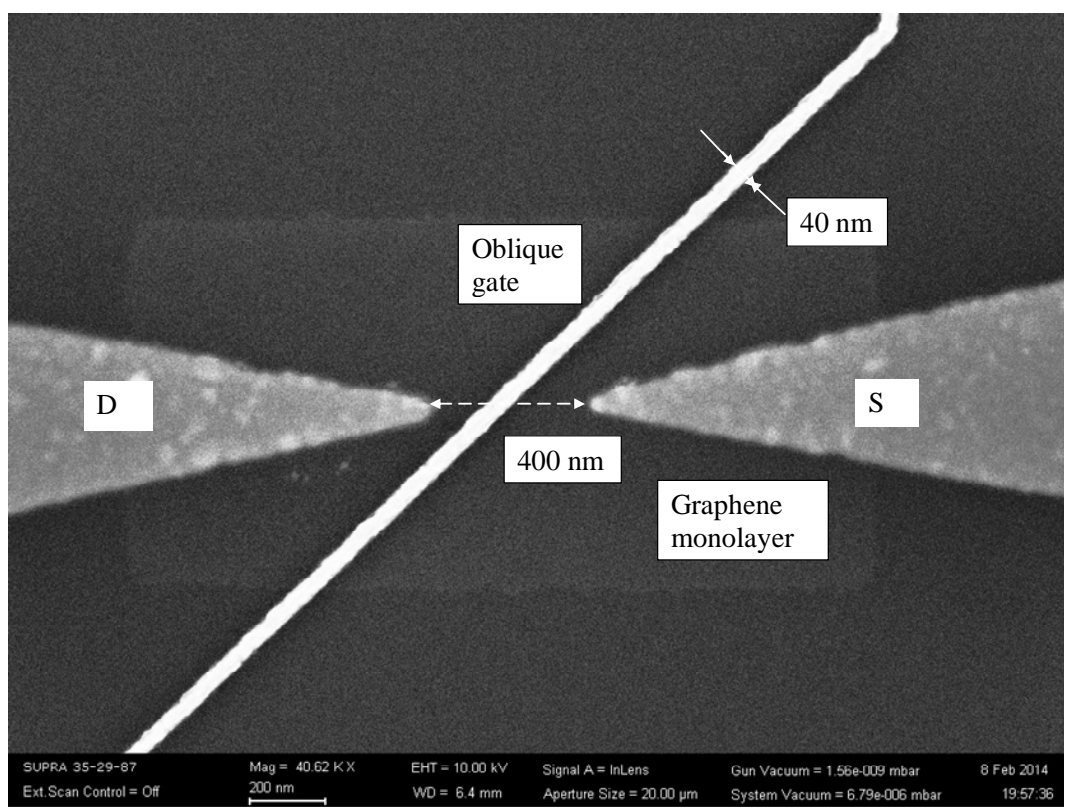

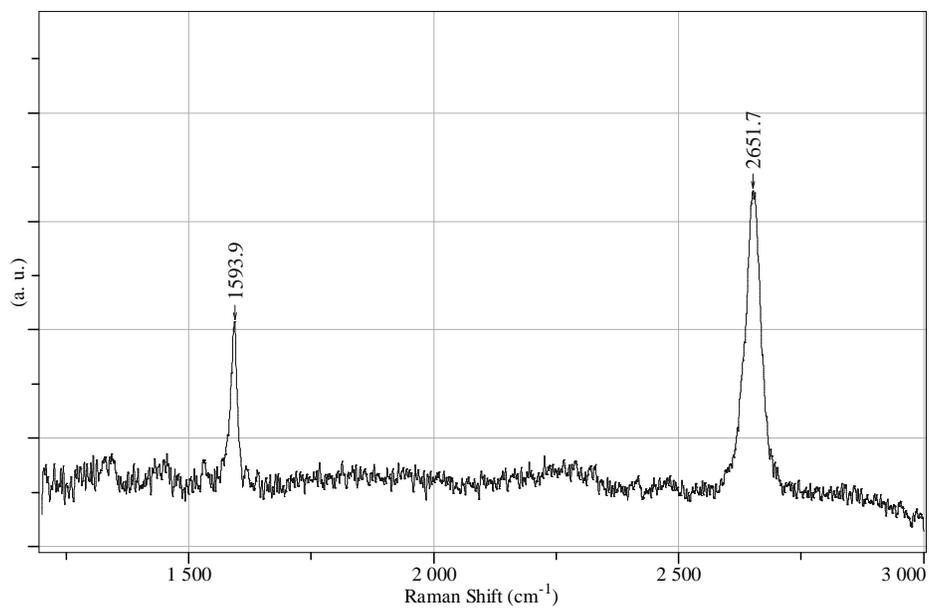

(a)

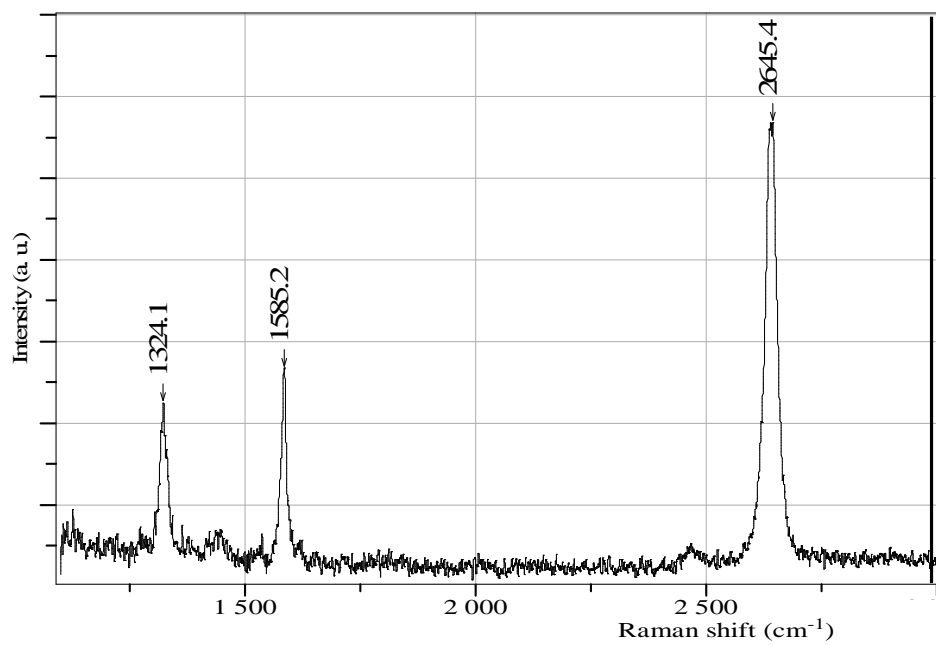

(b)

Fig. 2

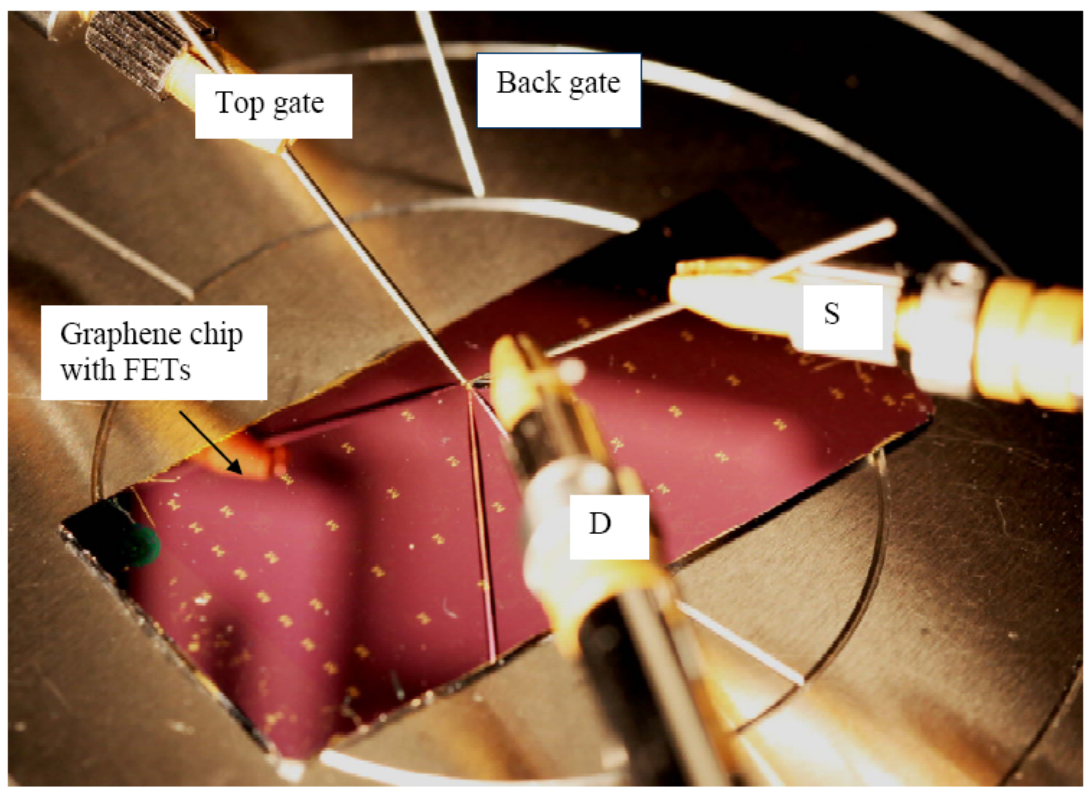

Fig. 3

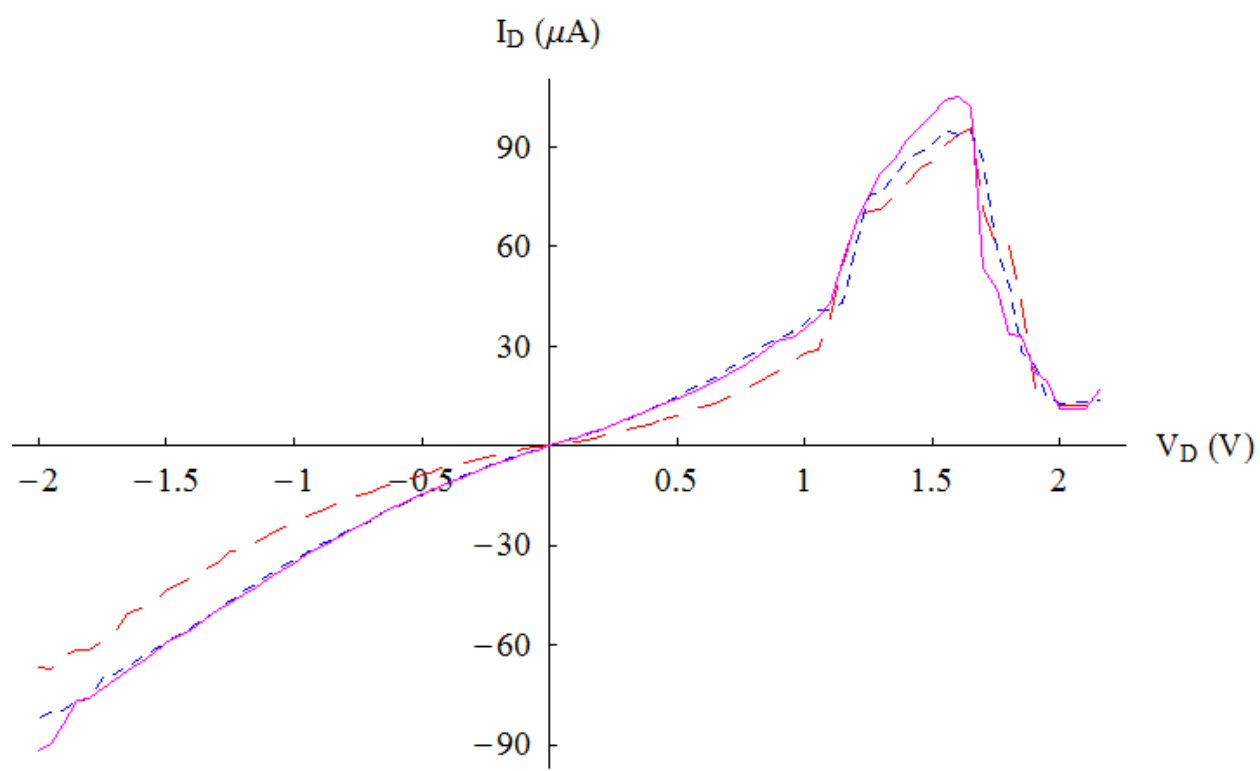

Fig. 4

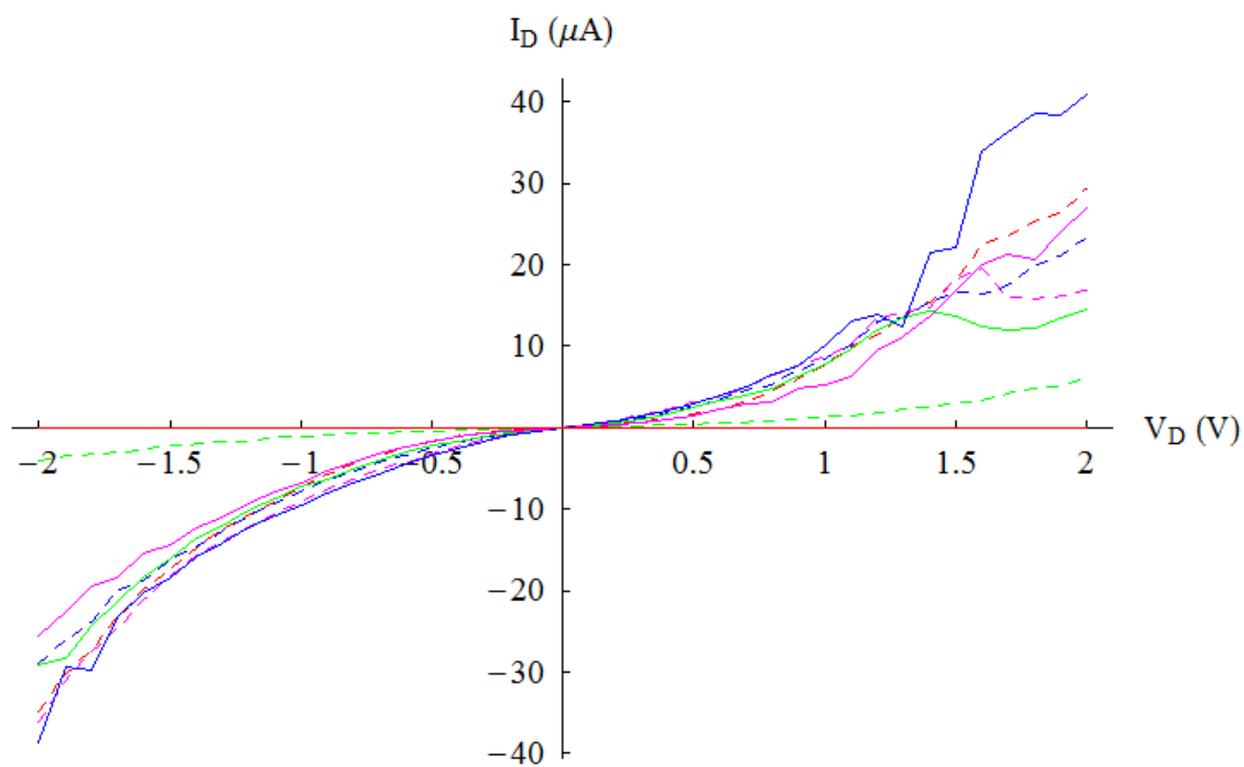

Fig. 5